\documentstyle[preprint,array,aps]{revtex}

\begin{document}
\draft
\title
{Entangled multiplet by angular momentum addition}
\author{Hyuk-jae Lee$^1${\footnote{e-mail:lhjae@iquips.uos.ac.kr}},
Sung Dahm Oh$^2${\footnote{e-mail:sdoh@sookmyung.ac.kr}} and
Doyeol Ahn$^{1,3}${\footnote{e-mail:dahn@uoscc.uos.ac.kr}}}
\address{
$^1$Institute of Quantum Information Processing and Systems,
University of Seoul, Seoul, 130-743, Korea\\
$^2$ Department of Physics, Sookmyung Women's University, Seoul,140-742, Korea\\
$^3$Department of Electrical and Computer Engineering,
University of Seoul,
Seoul, 130-743, Korea\\
}

%


\baselineskip24pt


\maketitle

\begin{abstract}
We present the visible entangled states of 4-qubit system which can be observed easily in physical laboratories.
This was motivated from the fact that the entangled state of 2-qubit system
comes from singlet and triplet states which are constructed through the angular momentum addition formalism.
We show that 4-qubit system has the new entangled states different from GHZ or W types entangled states.

\end{abstract}
\vspace{.25in}


\newpage

Entanglement has been in a crucial issue since the beginning of quantum mechanics.
In the new revolution of recent quantum theory, entanglement of quantum states is also of fundamental
interest for quantum information processing such as quantum
computation\cite{benn}--\cite{loss},
teleportation\cite{bras}--\cite{lee}, quantum key distribution\cite{eker} and clock
synchronization\cite{chua}--\cite{hwan}. Entanglement has been remained difficult
even though there were several investigations to qualify the entanglement and to classify types of the entangled states.
In the bipartite case, the pure state is the simplest in every respect.
However, as already known the multipartite situation is more complicated to classify the nature of the entanglement
and quantify the degree of entanglement. The mathematical and physical structure of entanglement is
not yet fully understood for multipartite case. Some groups suggested the way how to compare the entanglement properties of
pure multipartite qubits.\cite{meye}\cite{brie}

In order to understand the mathematical and physical structure of entanglement we need to recognize the visible entangled states.
The pure entangled state of bipartite case
is the Einstein-Podolsky-Rosen (EPR) state. The three particle system has the entangled states that can be classified
by two types; Greenberger-Horne-Zeilinger (GHZ) and W states. The four particle system also has GHZ and W states
as visible states. One believes the four particle system has
more states beyond GHZ and W states. The authors of Ref. \cite{brie} constructed the new type visible entangled
state, $|\phi_4\rangle$. Here we consider how to
construct the visible entangled states by angular momentum addition formalism. Then we try to find new entangled states of the four particles system.
Historically, the entangled states of two qubits were begun by spin-$\frac{1}{2}$ singlet state that
comes from the spin coupling. So we can suggest the entangled states of any multiqubit system through the angular momentum
addition formalism which is well known in the area of atomic and nuclear physics.

Let us start with the two spin-$\frac{1}{2}$ particles. We can combine two spin angular momenta
to form a total spin momentum $\vec{S}$,
\begin{equation}
\vec{S}=\vec{S_1}+\vec{S_2}.
\end{equation}
Then we have two classes of the basis of Hilbert space. The first class
corresponds to eigenstates of a complete set of commuting
operators, $\vec{S}^2_1, S_{1z}, \vec{S}^2_2, S_{2z}$. The
corresponding eigenstates will be written as
\begin{equation}
|s_1 m_1 s_2 m_2\rangle,
\end{equation}
where $s_i (i=1, 2)$ and $m_i$ denote the eigenvalues of $\vec{S}^2_i$ and $S_{iz}$, respectively, as
\begin{eqnarray}
\vec{S}^2_i |s_1 s_2 m_1 m_2\rangle&=&s_i (s_i + 1)\hbar^2|s_1 s_2 m_1 m_2\rangle\nonumber\\
S_{iz}|s_1 s_2 m_1 m_2\rangle&=&m_i\hbar|s_1 s_2 m_1 m_2\rangle.
\label{qstate}
\end{eqnarray}
If there is no coupling between two spins, the quantum state can be written by the tensor product
of eigenstate of each particle as
Eq. (\ref{qstate}).
The second class corresponds to the eigenstate of a complete set of commuting
operators, $\vec{S}^2_1, \vec{S}^2_2, \vec{S}^2, S_{z}$. The corresponding eigenstates will be written as
\begin{equation}
|s_1 m_1 s m\rangle,
\end{equation}
where $s$ and $m$ denote the eigenvalues of $\vec{S}^2_i$ and $S_{iz}$, respectively, as
\begin{eqnarray}
\vec{S}^2 |s_1 s_2 s m\rangle&=&s (s + 1)\hbar^2|s_1 s_2 s m \rangle\nonumber\\
S_{z}|s_1 s_2 s m\rangle&=&m\hbar|s_1 s_2 s m\rangle.
\label{qstatei}
\end{eqnarray}
There is the unitary transformation connecting these two basis called by Wigner or Clebsch-Gordan coefficient.
When an interaction between two spins is introduced as perturbation,
$\vec{S}^2$ and $S_z$ may be conserved, but not
the individual z-component $S_{1z}$ and $S_{2z}$. The state of Eq. (\ref{qstatei}) diagonalize the $\vec{S}^2$ and
$S_z$ simultaneously and can be basis vectors. The states of Eq. (\ref{qstatei}) can be written by the linear
combinations of the uncoupled states of Eq. (\ref{qstate})
as
\begin{eqnarray}
&&|\frac{1}{2} \frac{1}{2} 1 1\rangle= |\uparrow\uparrow\rangle,\\
&&|\frac{1}{2} \frac{1}{2} 1 0\rangle= \frac{1}{\sqrt{2}}(|\uparrow\downarrow\rangle+|\downarrow\uparrow\rangle),\label{trip}\\
&&|\frac{1}{2} \frac{1}{2} 1 -1\rangle= |\downarrow\downarrow\rangle,\\
&&|\frac{1}{2} \frac{1}{2} 0 0\rangle= \frac{1}{\sqrt{2}}(|\uparrow\downarrow\rangle-|\downarrow\uparrow\rangle)\label{sing},
\end{eqnarray}
where in right hand side $\uparrow$ and $\downarrow$ denote the spin up and the spin down, respectively.
Here we omit the eigenvalues, $s_1$ and $s_2$, for simplicity and wrote the eigenstate as $|m_1 m_2\rangle$.
The states in Eqs. (\ref{trip}) and (\ref{sing}) are two components of Bell basis which are maximally entangled.
The entangled states can be differentiated by eigenvalues of total spin and $z$-component.

We can also construct the entangled states of three spin-$\frac{1}{2}$ particles system by the same method. We have two kinds of
quantum states. The one is the state of an uncoupled representation and the other is the states of a coupled representation.
The coupled states are eigenstates of $\vec{S}_{12}^2=(\vec{S}_1 +\vec{S}_2)^2$, $\vec{S}^2=(\vec{S}_{12} + \vec{S}_3)^2$, $\vec{S}_3^2$ and
$S_z=S_{1z}+S_{2z}+S_{3z}$. After simple calculation, we can obtain the relation between the states of the uncoupled representations and
of the coupled representation as
\begin{mathletters}
\begin{eqnarray}
|\frac{1}{2}\frac{1}{2} 1 \frac{1}{2} \frac{3}{2} \frac{3}{2}\rangle&=&|\uparrow\uparrow\uparrow\rangle\label{thi}\\
|\frac{1}{2}\frac{1}{2} 1 \frac{1}{2} \frac{3}{2} \frac{1}{2}\rangle&=&\sqrt{\frac{1}{3}}(|\uparrow\downarrow\uparrow\rangle
+|\downarrow\uparrow\uparrow\rangle +|\uparrow\uparrow\downarrow\rangle)\label{thii}\\
|\frac{1}{2}\frac{1}{2} 1 \frac{1}{2} \frac{3}{2} -\frac{1}{2}\rangle&=&\sqrt{\frac{1}{3}}(|\downarrow\downarrow\uparrow\rangle
+|\uparrow\downarrow\downarrow\rangle +|\downarrow\uparrow\downarrow\rangle)\label{thiii}\\
|\frac{1}{2}\frac{1}{2} 1 \frac{1}{2} \frac{3}{2} -\frac{3}{2}\rangle&=&|\downarrow\downarrow\downarrow\rangle\label{thiv}\\
|\frac{1}{2}\frac{1}{2} 1 \frac{1}{2} \frac{1}{2} \frac{1}{2}\rangle&=&-\sqrt{\frac{1}{6}}(|\uparrow\downarrow\uparrow\rangle
+|\downarrow\uparrow\uparrow\rangle) +\sqrt{\frac{2}{3}}|\uparrow\uparrow\downarrow\rangle\label{thv}\\
|\frac{1}{2}\frac{1}{2} 1 \frac{1}{2} \frac{1}{2} -\frac{1}{2}\rangle&=&\sqrt{\frac{1}{6}}(|\uparrow\downarrow\downarrow\rangle
+|\downarrow\uparrow\downarrow\rangle) -\sqrt{\frac{2}{3}}|\downarrow\downarrow\uparrow\rangle\label{thvi}\\
|\frac{1}{2}\frac{1}{2} 0 \frac{1}{2} \frac{1}{2} \frac{1}{2}\rangle&=&\sqrt{\frac{1}{2}}(|\uparrow\downarrow \rangle-
|\downarrow\uparrow\rangle)\otimes|\uparrow \rangle\label{thvii}\\
|\frac{1}{2}\frac{1}{2} 0 \frac{1}{2} \frac{1}{2} -\frac{1}{2}\rangle&=&\sqrt{\frac{1}{2}}(|\uparrow\downarrow \rangle
-|\downarrow\uparrow\rangle)\otimes|\downarrow\rangle \label{thviii}.
\end{eqnarray}
\end{mathletters}
The states in Eqs. (\ref{thii}) and (\ref{thiii}) are W states. The states from Eq. (\ref{thi}) to Eq. (\ref{thviii}) do not
have GHZ states, GHZ state may be constructed by the linear combination of Eq. (\ref{thi}) and Eq. (\ref{thiv}). This implies
that the entangled state of tripartite system can be constructed by the spin coupling of three spin-$\frac{1}{2}$ particles.

Here we can have alternative coupled representations. The first one couples $\vec{S}_2$ and $\vec{S}_3$ to give $\vec{S}_{23}$ and add
$\vec{S}_1$ to give the total spin, $\vec{S}$. The other couples $\vec{S}_1$ and $\vec{S}_3$ to give $\vec{S}_{13}$ and add
$\vec{S}_2$ to give the total spin, $\vec{S}$.
These three coupled representations are  related to the coupling order and not independent because they are connected by
a linear transformation. The situation of coupling orders can be also happen in any multiqubit systems. Thus we will use the necessary
representations for our purpose.

When we have four spin momentum vectors, we may also use an uncoupled representation or one in which the vectors couple to a resultant
$\vec{S}$ and $M$, that is , an eigenstate of $\vec{S}^2=(\vec{S}_1+\vec{S}_2+\vec{S}_3+\vec{S}_4)^2$ and
$S_z=S_{1z}+S_{2z}+S_{3z}+S_{4z}$. However, the latter is no longer unique and requires further quantum numbers. There
are two possibilities for the further quantum numbers. The first case is an addition of two spin-1 vectors come
from resultant of two spin-$\frac{1}{2}$ vectors. The second case is an consecutive addition of four half spins.

At first, we consider a case that couples $\vec{S}_1$ and $\vec{S}_2$
to a resultant $\vec{S}_{12}$, $\vec{S}_3$ and $\vec{S}_4$ to $\vec{S}_{34}$ and finally adding $\vec{S}_{12}$ and
$\vec{S}_{34}$ to give $\vec{S}$. The final state can be written as
\begin{equation}
|(S_1 S_2)S_{12},(S_3 S_4)S_{34}, SM\rangle
\label{stat}
\end{equation}
being an common eigenstate of the spin operators
$\vec{S_1}^2, \vec{S_2}^2, \vec{S_3}^2, \vec{S_4}^2, \vec{S_{12}}^2,
\vec{S_{34}}^2, \vec{S}^2$ and $S_z$.
These form a complete set for describing its angular momentum properties.
This situation describes the coupling between four particles. The eigenstates of these spin operators
can be written as
the linear combinations of uncoupled states,
\begin{mathletters}
\begin{eqnarray}
&&|\frac{1}{2} 1 \frac{1}{2}122\rangle=|\uparrow\uparrow\uparrow\uparrow\rangle\\
&&|\frac{1}{2} 1 \frac{1}{2}121\rangle=\frac{1}{2}(|\uparrow\downarrow\uparrow\uparrow\rangle
+|\downarrow\uparrow\uparrow\uparrow\rangle+|\uparrow\uparrow\uparrow\downarrow\rangle
+|\uparrow\uparrow\downarrow\uparrow\rangle),\label{wstatei}\\
&&|\frac{1}{2} 1 \frac{1}{2}120\rangle=\frac{1}{\sqrt{6}}(|\downarrow\downarrow\uparrow\uparrow\rangle
+|\uparrow\downarrow\uparrow\downarrow\rangle+|\uparrow\downarrow\downarrow\uparrow\rangle
+|\downarrow\uparrow\uparrow\downarrow\rangle
+|\downarrow\uparrow\downarrow\uparrow\rangle+|\uparrow\uparrow\downarrow\downarrow\rangle),\label{unkn}\\
&&|\frac{1}{2} 1 \frac{1}{2}12-1\rangle=\frac{1}{2}(|\downarrow\downarrow\uparrow\downarrow\rangle
+|\downarrow\downarrow\downarrow\uparrow\rangle+|\uparrow\downarrow\downarrow\downarrow\rangle
+|\downarrow\uparrow\downarrow\downarrow\rangle),\label{wstateii}\\
&&|\frac{1}{2} 1 \frac{1}{2}122\rangle=|\downarrow\downarrow\downarrow\downarrow\rangle,\\
&&|\frac{1}{2} 1 \frac{1}{2}111\rangle=\frac{1}{2}(|\uparrow\downarrow\uparrow\uparrow\rangle
+|\downarrow\uparrow\uparrow\uparrow\rangle-|\uparrow\uparrow\uparrow\downarrow\rangle
-|\uparrow\uparrow\downarrow\uparrow\rangle),\label{wstateiii}\\
&&|\frac{1}{2} 1 \frac{1}{2}110\rangle=\frac{1}{\sqrt{2}}(|\downarrow\downarrow\uparrow\uparrow\rangle
-|\uparrow\uparrow\downarrow\downarrow\rangle),\label{ghz}\\
&&|\frac{1}{2} 1 \frac{1}{2}11-1\rangle=\frac{1}{2}(|\downarrow\downarrow\uparrow\downarrow\rangle
+|\downarrow\downarrow\downarrow\uparrow\rangle-|\uparrow\downarrow\downarrow\downarrow\rangle
-|\downarrow\uparrow\downarrow\downarrow\rangle),\label{wstateiv}\\
&&|\frac{1}{2} 1 \frac{1}{2}100\rangle=\frac{1}{\sqrt{3}}(|\uparrow\uparrow\downarrow\downarrow\rangle
+|\downarrow\downarrow\uparrow\uparrow\rangle)-\frac{1}{2\sqrt{3}}(|\uparrow\downarrow\uparrow\downarrow\rangle
+|\uparrow\downarrow\downarrow\uparrow\rangle
+|\downarrow\uparrow\uparrow\downarrow\rangle+|\downarrow\uparrow\uparrow\downarrow\rangle),\label{unknii}\\
&&|\frac{1}{2} 0 \frac{1}{2}111\rangle=\frac{1}{\sqrt{2}}(|\uparrow\downarrow\rangle
-|\downarrow\uparrow\rangle)\otimes|\uparrow\uparrow\rangle,\\
&&|\frac{1}{2} 0 \frac{1}{2}110\rangle=\frac{1}{2}(|\uparrow\downarrow\rangle
-|\downarrow\uparrow\rangle)\otimes(|\uparrow\downarrow\rangle+|\downarrow\uparrow\rangle),\\
&&|\frac{1}{2} 0 \frac{1}{2}11-1\rangle=\frac{1}{\sqrt{2}}(|\uparrow\downarrow\rangle
-|\downarrow\uparrow\rangle)\otimes|\downarrow\downarrow\rangle\\
&&|\frac{1}{2} 1 \frac{1}{2}011\rangle=\frac{1}{\sqrt{2}}|\uparrow\uparrow\rangle\otimes(|\uparrow\downarrow\rangle
-|\downarrow\uparrow\rangle),\\
&&|\frac{1}{2} 1 \frac{1}{2}010\rangle=\frac{1}{2}(|\uparrow\downarrow\rangle
+|\downarrow\uparrow\rangle)\otimes(|\uparrow\downarrow\rangle-|\downarrow\uparrow\rangle),\\
&&|\frac{1}{2} 1 \frac{1}{2}01-1\rangle=\frac{1}{\sqrt{2}}|\downarrow\downarrow\rangle\otimes(|\uparrow\downarrow\rangle
-|\downarrow\uparrow\rangle),\\
&&|\frac{1}{2} 0 \frac{1}{2}000\rangle=\frac{1}{2}(|\uparrow\downarrow\rangle
-|\downarrow\uparrow\rangle)\otimes(|\uparrow\downarrow\rangle-|\downarrow\uparrow\rangle),
\end{eqnarray}
\end{mathletters}
where the left hand sides is basis of coupled case and the right
hand the linear combinations of an uncoupled representation. The
above basis vectors give the entangled states as 4-GHZ in Eq.
(\ref{ghz}), W states in Eqs. (\ref{wstatei}), (\ref{wstateii}),
(\ref{wstateiii}) and (\ref{wstateiv}) and the unknown entangled
state in Eqs. (\ref{unkn}).  The Eq. (\ref{unknii}) is similar to Eqs. (\ref{unkn}) except with different coefficients.
The new entangled state in Eq. (\ref{unkn}) is distinguished
from GHZ and W states. However, this unknown states is similar to
the W state. If one qubit among the W states is traced out then the
remaining qubits are entangled and are a three qubit W
state by means of filtering measurement. However if one qubit
among Eq. (\ref{unkn}) is traced out then the remaining qubits
are entangled in a three qubit W states with
certainty.

The entanglement measures for the
pure states in multipartite system are suggested by some groups. The authors in Ref.
\cite{meye} defined the global entanglement in multiparticle
systems. They defined Q function from states in Hilbert space to
nonnegative real number. We can calculate the value of Q function for our entangled
states in Eqs. (12). The GHZ state and the state in Eq. (\ref{unkn}) are $1$ and the W state is $\frac{3}{4}$.
Then the Q function does not enable to distinguish the GHZ states from the state in Eq. (\ref{unkn}).
The authors in Ref. \cite{brie} introduced the
maximal connectedness and the persistency of entanglement of multiqubit system.
According to their definition the GHZ state is maximally
connected but W state not. However, their persistency are $P_e =1$
and $P_e=3$ for the GHZ and W state , respectively. The state
in Eq. (\ref{unkn}) is not maximally
connected and the persistency is $P_e=3$. These facts indicate us that
maximal connectedness and persistency can not distinguish the W
state from the state in Eq. (\ref{unkn}).

We can have the second possibility with a consecutive additions of four spin-$\frac{1}{2}$ system. Then we choose
other quantum numbers as $S_1, S_{12}, S_{123}, S, M$.  We obtain eigenstates
which specifies the state of the coupled four vector system:
\begin{mathletters}
\begin{eqnarray}
&&|\frac{1}{2} \frac{1}{2} 1 \frac{1}{2}\frac{3}{2}\frac{1}{2}22\rangle=|\uparrow\uparrow\uparrow\uparrow\rangle\\
&&|\frac{1}{2} \frac{1}{2} 1 \frac{1}{2}\frac{3}{2}\frac{1}{2}21\rangle=
\frac{1}{2}(|\uparrow\downarrow\uparrow\uparrow\rangle+|\downarrow\uparrow\uparrow\uparrow\rangle
+|\uparrow\uparrow\uparrow\downarrow\rangle+|\uparrow\uparrow\downarrow\uparrow\rangle),\label{iwstatei}\\
&&|\frac{1}{2} \frac{1}{2} 1 \frac{1}{2}\frac{3}{2}\frac{1}{2}20\rangle=
\frac{1}{\sqrt{6}}(|\downarrow\downarrow\uparrow\uparrow\rangle+|\uparrow\downarrow\uparrow\downarrow\rangle
+|\uparrow\downarrow\downarrow\uparrow\rangle+|\downarrow\uparrow\uparrow\downarrow\rangle
+|\downarrow\uparrow\downarrow\uparrow\rangle+|\uparrow\uparrow\downarrow\downarrow\rangle),\label{iunkn}\\
&&|\frac{1}{2} \frac{1}{2} 1 \frac{1}{2}\frac{3}{2}\frac{1}{2}2-1\rangle=
\frac{1}{2}(|\downarrow\downarrow\uparrow\downarrow\rangle
+|\downarrow\downarrow\downarrow\uparrow\rangle+|\uparrow\downarrow\downarrow\downarrow\rangle
+|\downarrow\uparrow\downarrow\downarrow\rangle),\label{iwstateii}\\
&&|\frac{1}{2} \frac{1}{2} 1 \frac{1}{2}\frac{3}{2}\frac{1}{2}2-2\rangle=
|\downarrow\downarrow\downarrow\downarrow\rangle,\\
&&|\frac{1}{2} \frac{1}{2} 1 \frac{1}{2}\frac{3}{2}\frac{1}{2}11\rangle=
-\frac{1}{\sqrt{12}}(|\uparrow\downarrow\uparrow\uparrow\rangle
+|\downarrow\uparrow\uparrow\uparrow\rangle+|\uparrow\uparrow\downarrow\uparrow\rangle)
+\frac{\sqrt{3}}{2}|\uparrow\uparrow\uparrow\downarrow\rangle),\label{iunknii}\\
&&|\frac{1}{2} \frac{1}{2} 1 \frac{1}{2}\frac{3}{2}\frac{1}{2}10\rangle=
\frac{1}{\sqrt{6}}(|\uparrow\uparrow\downarrow\downarrow\rangle
+|\uparrow\downarrow\uparrow\downarrow\rangle+|\downarrow\uparrow\uparrow\downarrow\rangle
-|\downarrow\downarrow\uparrow\uparrow\rangle
-|\uparrow\downarrow\downarrow\uparrow\rangle-|\downarrow\uparrow\downarrow\uparrow\rangle),\label{iunkn}\\
&&|\frac{1}{2} \frac{1}{2} 1 \frac{1}{2}\frac{3}{2}\frac{1}{2}1-1\rangle=
-\frac{\sqrt{3}}{2}|\downarrow\downarrow\downarrow\uparrow\rangle+\frac{1}{\sqrt{12}}
(|\downarrow\downarrow\uparrow\downarrow\rangle+|\uparrow\downarrow\downarrow\downarrow\rangle
+|\downarrow\uparrow\downarrow\downarrow\rangle),\label{inunkniv}\\
&&|\frac{1}{2} \frac{1}{2} 1 \frac{1}{2}\frac{1}{2}\frac{1}{2}11\rangle=
\frac{1}{\sqrt{3}}(|\uparrow\uparrow\downarrow\downarrow\rangle+|\downarrow\downarrow\uparrow\uparrow\rangle)
-\frac{1}{2\sqrt{3}}(|\uparrow\downarrow\uparrow\downarrow\rangle+|\uparrow\downarrow\downarrow\uparrow\rangle
+|\downarrow\uparrow\uparrow\downarrow\rangle+|\downarrow\uparrow\uparrow\downarrow\rangle),\label{iunknii}\\
&&|\frac{1}{2} \frac{1}{2} 1 \frac{1}{2}\frac{1}{2}\frac{1}{2}10\rangle=
\frac{1}{\sqrt{2}}(|\uparrow\downarrow\rangle-|\downarrow\uparrow\rangle)\otimes|\uparrow\uparrow\rangle,\\
&&|\frac{1}{2} \frac{1}{2} 1 \frac{1}{2}\frac{1}{2}\frac{1}{2}1-1\rangle=
\frac{1}{2}(|\uparrow\downarrow\rangle-|\downarrow\uparrow\rangle)\otimes(|\uparrow\downarrow\rangle
+|\downarrow\uparrow\rangle),\\
&&|\frac{1}{2} \frac{1}{2} 1 \frac{1}{2}\frac{1}{2}\frac{1}{2}00\rangle=
\frac{1}{\sqrt{2}}(|\uparrow\downarrow\rangle-|\downarrow\uparrow\rangle)\otimes|\downarrow\downarrow\rangle\\
&&|\frac{1}{2} \frac{1}{2} 0 \frac{1}{2}\frac{1}{2}\frac{1}{2}11\rangle=
\frac{1}{\sqrt{2}}(|\uparrow\downarrow\rangle-|\downarrow\uparrow\rangle)\otimes|\uparrow\uparrow\rangle,\\
&&|\frac{1}{2} \frac{1}{2} 0 \frac{1}{2}\frac{1}{2}\frac{1}{2}10\rangle=
\frac{1}{2}(|\uparrow\downarrow\rangle-|\downarrow\uparrow\rangle)\otimes(|\uparrow\downarrow\rangle
+|\downarrow\uparrow\rangle),\\
&&|\frac{1}{2} \frac{1}{2} 0 \frac{1}{2}\frac{1}{2}\frac{1}{2}1-1\rangle=
\frac{1}{\sqrt{2}}(|\uparrow\downarrow\rangle-|\downarrow\uparrow\rangle)\otimes|\downarrow\downarrow\rangle,\\
&&|\frac{1}{2} \frac{1}{2} 0 \frac{1}{2}\frac{1}{2}\frac{1}{2}00\rangle=
\frac{1}{2}(|\uparrow\downarrow\rangle-|\downarrow\uparrow\rangle)\otimes(|\uparrow\downarrow\rangle
-|\downarrow\uparrow\rangle).
\end{eqnarray}
\end{mathletters}
These states have the W state and new entangled states like Eqs. (\ref{iunkn}) and (\ref{iunknii})
but do not have the GHZ state. As three particle case, the GHZ state can be constructed by linear combination.

This letter suggests the visible entangled states in four particle
system. These states in Eqs. (12) and Eqs. (13) contain GHZ
and W type entangled states and other types which may play important
roles to check the mathematical and physical structure of
entanglement. For example, we check the measurements of
entanglements  which distinguish well in pure tripartite systems  are not able in
the four particle systems.  This fact shows that we need the new
entanglement measure to compare the entanglement properties for
multipartite qubit systems.  The
generalization to higher multipartite case is straightforward
and we believe that our visible entangled states
will play a role to find the new entanglement measure. We believe that
the physical system of these entanglement construction scenario can be easily
implemented in the experiment.

\vspace{2.0cm}

\centerline{\bf Acknowledgements}

H. -j Lee and D. Ahn were supported by the Korean Ministry of
Science and Technology through the Creative Research Initiatives
Program under Contact No. M10116000008-02F0000-00610  and S. D. Oh
was supported by the Korea Science and Engineering Foundation
through the Contract No. R06-2002-007-01003-0(2002) and by Korea
Research Foundation through the contract No. KRF-2002-070-c00029.

\end{document}